\documentclass{article}
\usepackage{spconf,amsmath,graphicx}

\usepackage{enumitem}
\setlist{nosep, leftmargin=14pt}

\usepackage{mwe} 
\usepackage{booktabs}
\usepackage[table,xcdraw]{xcolor}

\title{Homodyned K-distribution: parameter estimation and uncertainty quantification using Bayesian neural networks}
\name{Ali K. Z. Tehrani$^{1}$, Ivan M. Rosado-Mendez$^{2}$, and Hassan Rivaz$^{1}$\thanks{e-mail: A\_Kafaei@encs.concordia.ca, rosadomendez@wisc.edu, and hrivaz@ece.concordia.ca}}
\address{$^{1}$Department of Electrical and Computer Engineering, Concordia University, Canada.\\	
$^{2}$Department of Medical Physics, and Radiology, University of Wisconsin, United States.}

\begin{document}
%
\maketitle
\begin{abstract}
Quantitative ultrasound (QUS) allows estimating the intrinsic tissue properties. Speckle statistics are the QUS parameters that describe the first order statistics of ultrasound (US) envelope data. The parameters of Homodyned K-distribution (HK-distribution) are the speckle statistics that can model the envelope data in diverse scattering conditions. However, they require a large amount of data to be estimated reliably. Consequently, finding out the intrinsic uncertainty of the estimated parameters can help us to have a better understanding of the estimated parameters. In this paper, we propose a Bayesian Neural Network (BNN) to estimate the parameters of HK-distribution and quantify the uncertainty of the estimator.          
\end{abstract}
\begin{keywords}
Quantitative ultrasound, Tissue characterization, Homodyned K-distribution, Uncertainty, Bayesian Neural Network
\end{keywords}
\section{Introduction}
\label{sec:intro}

Quantitative ultrasound (QUS) aims to characterize the tissue by revealing information about the scatterers. These microstructures are smaller than the wavelength and scatter the ultrasound wave. Speckle statistics provide insight about the number and coherency of the scatterers which are correlated with the tissue properties \cite{wagner1983statistics,oelze2016review}. The scatterer density is an important property of the tissue which is defined as the number of scatterers per resolution cell (an ellipsoidal volume defined by - 6 dB  point  of  the  beam  profile \cite{wagner1983statistics}). Coherency of the scatterers is also another parameter that is related to spatial organization of the scatterers. Homodyned K-distribution can comprehensively model the envelope data under diverse number of scatterers (from low to high) and coherency levels.

The HK-distribution does not have closed form solution and conventional methods of estimating the values of the parameters of the HK distribution such as the method based on moments \cite{Hruska2009} and log compressed moments (we refer to it as XU) \cite{destrempes2013estimation}, rely on iterative optimization methods. In our previous works, we employed Convolutional Neural Networks (CNN) to classify and segment the US data into fully developed (high scatterer number density) and underdeveloped (low scatterer number density) speckle \cite{tehrani2021ultrasound,tehrani2022robust}. Recently, an Artificial Neural Network (ANN) was introduced by Zhou \textit{et al.} \cite{zhou2021parameter}. The proposed method was a Multi Layered Perceptron (MLP) that employed speckle statistics to estimate the parameters of HK-distribution.

The ANN estimator employs MLP layers which are prone to overfitting. In addition to this, there is no metric to investigate the reliability of the estimated value. In this paper, we address these two issues and aim to improve the estimation of the HK-distribution parameters and quantify the uncertainty using Bayesian Neural Networks (BNN). The proposed method can also be used to extract QUS parametric images, and detect the regions with high uncertainty.    
\section{Material and Method}
\subsection{Homodyned K-distribution parameters and training data generation}
The Homodyned K-distribution (HK-distribution) is defined as \cite{destrempes2013estimation}:
\begin{equation}
	P_{HK}(A|\varepsilon ,\sigma^2,\alpha) = A\int_{0}^{\infty}uJ_{0}(u\epsilon)J_{0}(uA)(1+\frac{u^2\sigma^2}{2})^{-\alpha}du
\end{equation}
where $\alpha$ is the scatterer clustering parameter that depends on the scatterer number density, $A$ is the envelope of the backscattered echo signal, and $J_{0}(.)$ denotes the zero-order Bessel function. The coherent signal power is $\epsilon^2$, and the diffuse signal power can be obtained by $2\sigma^2\alpha$ \cite{destrempes2013estimation}. The parameter $k$ is defined as the ratio of coherent to diffuse signal power and along with $\alpha$ has been employed widely for tissue characterization and we refer to them as HK-distribution parameters. The main purpose of this paper is to estimate $k$ and $log_{10}(\alpha)$ (similar to \cite{zhou2021parameter}) and quantify the uncertainty of their estimation.

In order to generate training data for both ANN and BNN, sampling from HK-distribution is required. Similar to \cite{Hruska2009,zhou2021parameter}, we employed the following equation produce synthetic samples from HK-distribution. 
	\begin{equation}
	a_i = \sqrt{\left (  \varepsilon +X\sigma \sqrt{Z/\alpha }\right )^2+\left (  Y\sigma \sqrt{Z/\alpha }\right )^2}
\end{equation}
where $X$ and $Y$ are independent and identically distributed (i.i.d) samples from unit Normal distribution, $a_i$ is the generated sample from HK-distribution, and $Z$ is sampled from the Gamma distribution with shape parameter $\alpha$ and scale parameter of 1.
To generate training data, $log_{10}(\alpha)$ is randomly selected from values ranging -0.3 to 1.4 which corresponds to $\alpha$ of 0.5 to 25. $k$ is also randomly selected from values ranging 0 to 1.

Different sizes of data, results in different values for the calculated feature. We generated different sizes of data (we refer to it as $N_s$) to train the networks (similar to \cite{zhou2021parameter}). The network is trained for each size separately using 10000 generated training data. The test data is generated with the same range of parameters, on total 31 and 11 distinct $log_{10}(\alpha)$ and $k$ values, respectively. For each value of $log_{10}(\alpha)$ and $k$, 100 test sets are generated; therefore there are $31\times 11\times100$ samples of test data for each $N_s$. 

\subsection{ANN estimator}
In \cite{zhou2021parameter}, Zhou \textit{et al.} proposed an ANN approach to estimated HK-distribution parameters and out-performed the XU optimization method \cite{destrempes2013estimation}. The procedure was as follows. First, SNR, skewness, Kurtosis, X and U statistics were computed. The equations to compute the parameters are given as: 
\begin{equation}
\begin{gathered}
		R_v=\frac{\overline{A^v}}{\sqrt{\overline{A^{2v}}-(\overline{A^v})^2}},\\ 
		S_v= \frac{\overline{(A^v-\overline{A^v})^3}}{(\overline{A^{2v}}-(\overline{A^v})^2)^{1.5}},\\
		K_v = \frac{\overline{A^{4v}}-4\overline{A^v} \times \overline{A^{3v}} + 6\overline{A^{2v}} \times \overline{A^{v}}^2-3\overline{A^{v}}^4}{(\overline{A^{2v}} -\overline{A^{v}}^2)^2},\\ 
		U = \overline{log(I)} - log(\overline{I}), \\
		X = \overline{I\times log(I)}/\overline{I}-\overline{log(I)},
\end{gathered}
\end{equation}
where $A$ is the envelope data, $I$ is the intensity ($I = A^2$), and $v$ is $\{0.72,0.88\}$ as suggested by Hruska \textit{et al.} \cite{Hruska2009} and Gao \textit{et al.} \cite{gao2021ultrasonic}. 

In the next step, $R_{0.88}$, $R_{0.72}$, $S_{0.88}$, $S_{0.72}$, $K_{0.88}$, $K_{0.72}$, $X$, and $U$ were employed as inputs of a MLP (ANN) as suggested by \cite{gao2021ultrasonic} to train the network which estimated the $log_{10}(\alpha)$ and $k$. We implemented this method for comparison and used the same network architecture (2 hidden layers with 10 and 4 nodes). We refer to this method as $ANN$.  
\subsection{Bayesian Neural Network (BNN)}
Let $Y$, $W$, and $X$ be the target, weights and input vectors, respectively. Assuming the training data be $D = \left \{ X_i,Y_i \right \}$, training a NN can be defined as:
\begin{equation}
	W^* = \arg \max_{W}\left \{ P(D|W) \right \}
\end{equation}   
where the optimum weights ($W^*$) are learned during the training and used in the test to predict $Y$. In BNN, the weights of the neural network are not fixed, and each weight is sampled from a distribution. During the training instead of learning the weights, the parameters of the distribution, from which the weights are sampled, are learned. Predicting $Y$ can be formulated as \cite{jospin2022hands}:

\begin{equation}
	P(Y|D) = \int_{W}P(Y|W)P(W|D)
\end{equation} 
where $p(W|D)$ is the posterior distribution of the weights which is learned during the training. The integration over all possible values of $W$ is intractable and computationally expensive. To resolve this issue, the posterior distribution $P(W|D)$ is sampled and the prediction $\tilde{Y}$ can be obtained by:
 \begin{equation}
 	\label{eq:sampling}
 	\begin{gathered}
\tilde{Y}\simeq \frac{1}{N}\sum_{i=1}^{N}\tilde{Y_i},\\
W_i \sim P(W|D),\\
\tilde{Y_i}\sim P(Y|W_i),
\end{gathered}
\end{equation}
where the operator $\sim$ denotes sampling from the distribution. Eq \ref{eq:sampling} can be simply explained as running the trained network multiple times (each forward pass of the network gives $\tilde{Y_i}$) and computing the mean value of predictions as the final estimated value. Uncertainty can also be quantified as the standard deviation of the predictions $\tilde{Y_i}$ which can be written as:
 \begin{equation}
	\label{eq:unc}
	\begin{gathered} 
uncertainty = \sqrt{Var(Y_i)},
\end{gathered}
\end{equation}
The Mean Absolute Error (MAE) loss is utilized for training which is also sampled multiple times by forwarding the inputs and sampling from the weights multiple times (here 6) to have a better approximation of the loss value. Two Bayesian hidden layers having 64 and 200 nodes with leaky Relu activation functions were employed and Adam optimizer is utilized for optimization.  
 
\begin{table}[]
	\caption{RRMSE and MAE of $log_{10}(\alpha)$ using different numbers of HK-distribution samples ($N_s$).}
	\label{tab:alpha}
	\centering
	\resizebox{0.33\textwidth}{!}{
		\centering
	\begin{tabular}{@{}ccccc@{}}
		\toprule
		& \multicolumn{2}{c}{ANN}       & \multicolumn{2}{c}{BNN}       \\ \midrule
		& \textbf{RRMSE} & \textbf{MAE} & \textbf{RRMSE} & \textbf{MAE} \\
		\rowcolor[HTML]{C0C0C0} 
		$N_s$ = 65536 & 0.054          & 0.048        & \textbf{0.012}          & \textbf{0.035}        \\
		\rowcolor[HTML]{FFFFFF} 
		$N_s$ = 16384 & 0.052          & 0.061        & \textbf{0.029}          & \textbf{0.054}        \\
		\rowcolor[HTML]{C0C0C0} 
		$N_s$ = 4096  & 0.125          & 0.091        & \textbf{0.090}          & \textbf{0.083}        \\
		\rowcolor[HTML]{FFFFFF} 
		$N_s$ = 1024  & 0.393          & 0.129        & \textbf{0.388}          & \textbf{0.123}        \\ \bottomrule
	\end{tabular}}
\end{table}
\begin{table}[]
	\caption{RRMSE and MAE of $k$ using different numbers of HK-distribution samples ($N_s$).}
	\label{tab:k}
	\centering
	\resizebox{0.33\textwidth}{!}{		
	\begin{tabular}{@{}ccccc@{}}
		\toprule
		& \multicolumn{2}{c}{ANN}       & \multicolumn{2}{c}{BNN}       \\ \midrule
		& \textbf{RRMSE} & \textbf{MAE} & \textbf{RRMSE} & \textbf{MAE} \\
		\rowcolor[HTML]{C0C0C0} 
		$N_s$ = 65536 & 0.143          & 0.074        & \textbf{0.122}          & \textbf{0.053}        \\
		\rowcolor[HTML]{FFFFFF} 
		$N_s$ = 16384 & \textbf{0.218}          & 0.084        & 0.235          & \textbf{0.073}        \\
		\rowcolor[HTML]{C0C0C0} 
		$N_s$ = 4096  & 0.359          & 0.118        & \textbf{0.291}          & \textbf{0.103}        \\
		\rowcolor[HTML]{FFFFFF} 
		$N_s$ = 1024  & 0.538          & 0.153        & \textbf{0.460}          & \textbf{0.139}        \\ \bottomrule
	\end{tabular}}
\end{table}

\begin{figure}	
	
	\centering
	\includegraphics[width=0.47\textwidth]{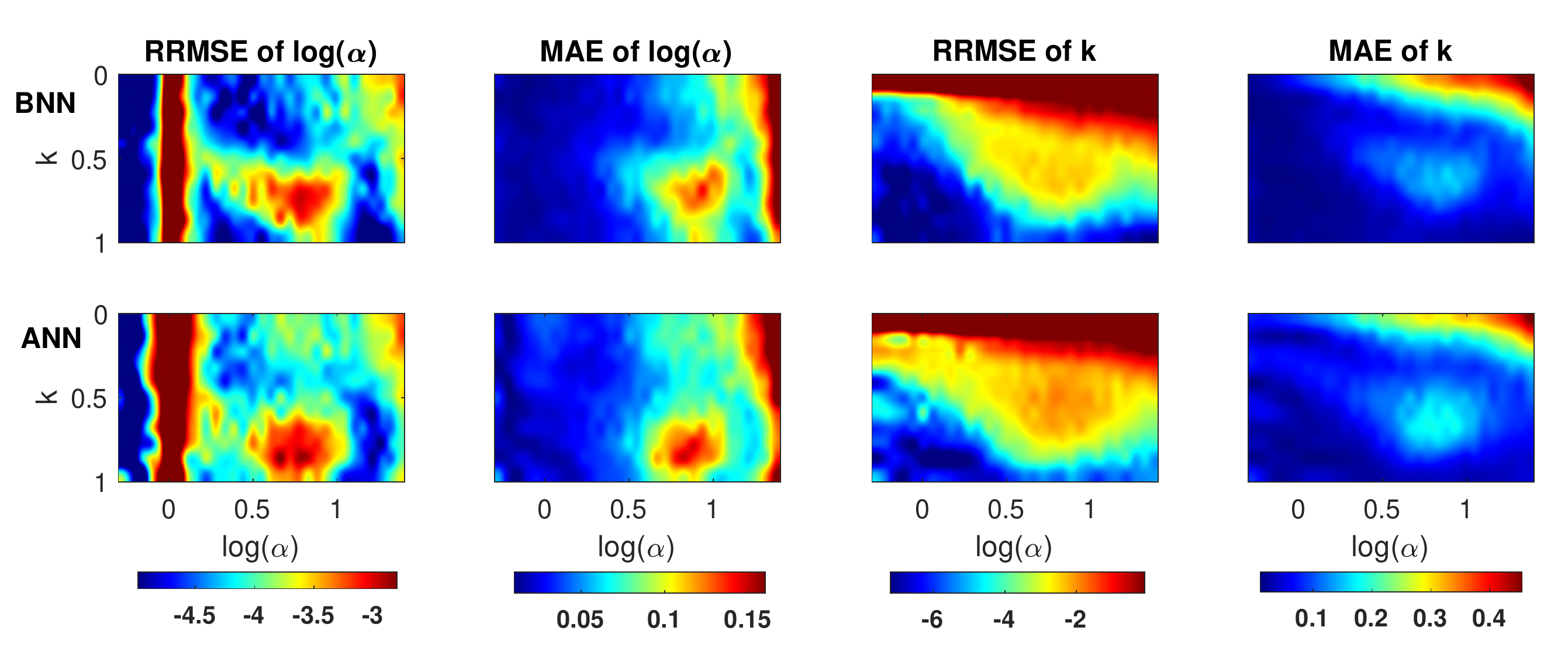}
	\caption{The RRMSE and MAE error maps of BNN (top) and ANN (bottom) for $N_s = 16384$. The RRMSEs are shown in log scale for better visualization.}
\label{fig:error_map}
\end{figure}
\section{Results}
\label{sec:res}
\subsection{Simulation Results}
The accuracy of the estimators of HK-distribution parameters heavily depends on the number of available i.i.d samples. We evaluated the methods using different number of samples ($N_s$). The Relative Root Mean Square Error (RRMSE) and MAE are employed as the metrics which can be defined as \cite{Hruska2009,zhou2021parameter}:
\begin{equation}
	\begin{gathered} 
	RRMSE = \sqrt{\frac{<(y-\widetilde{y})^{2}>}{|y|+\varepsilon }}, \\
	MAE = <|y-\widetilde{y}|>,
\end{gathered}
\end{equation}
 where ${<.}>$ denote averaging operation and $\varepsilon$ is a small number (here 0.001) to avoid division by zero. The simulation results for $log_{10}(\alpha)$ and $k$ are given in Tables \ref{tab:alpha} and \ref{tab:k}, respectively. According to the tables, the proposed BNN has lower error compared to ANN for estimation of both $log_{10}(\alpha)$ and $k$ in the most of sample sizes.

The RRMSE and MAE error maps are shown for $N_s=16384$ and different ground truth values of $log_{10}(\alpha)$ and $k$ . RRMSEs high values around the ground truth zero are due to the division by the small number. For better visualization, RRMSEs are plotted in log scale. Fig. \ref{fig:error_map} shows that the proposed BNN method has lower error than ANN (notice the blue regions in RRMSEs). 

\begin{figure}	
	
	\centering
	\includegraphics[width=0.38\textwidth]{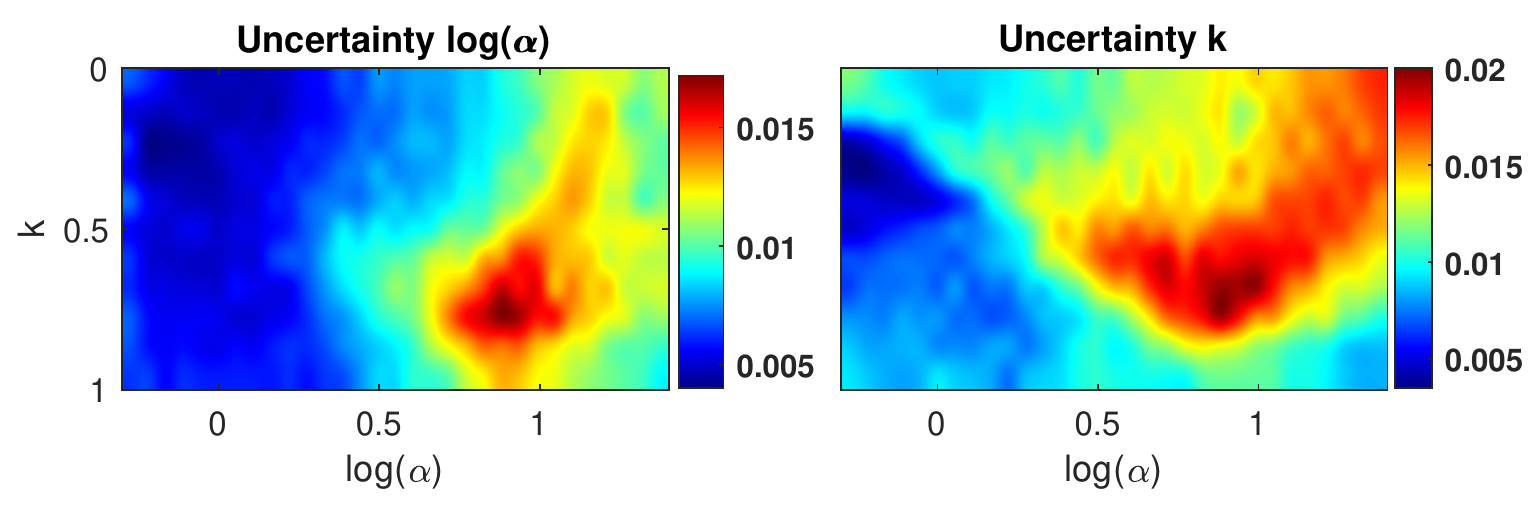}
	\caption{The estimated uncertainty (standard deviation of predictions) of $log_{10}(\alpha)$ and $k$ for $N_s = 16384$ using BNN. The areas with high uncertainty correspond to areas with high error in Fig. \ref{fig:error_map}.}
	\label{fig:uncertainty}
\end{figure}

The proposed method can also provide uncertainty of the prediction (Eq \ref{eq:unc}). Fig. \ref{fig:uncertainty} shows the uncertainty of the estimation of the parameters. It can be seen that areas in Fig. \ref{fig:error_map} that high error is presents, the uncertainty is high which can provide an insight about the reliability of the estimation.  

\subsection{Experimental Phantom Results}
\label{sec:majhead}
A two layered phantom was constructed from an emulsion of ultrafiltered milk and water-based gelatin having 5–43 $\mu m$ diameter glass beads (3000E, Potters Industries, Valley Forge, PA,  USA) as the source of scattering. Data was collected by a 18L6 probe, linear array transducer, using a Siemens Acuson S2000 scanner (Siemens Medical Solutions USA, Inc.) with operating center frequency of 8.9 MHz. The middle layer was made to have a higher backscattering coefficient than the other two layers by increasing the concentration of scatterers (higher $\alpha$). The backscattering coefficient of top and bottom layers is $3.52 \times 10^{-3}$  $cm^{-1}$ $sr^{-1}$ and it is $6.37 \times 10^{-3}$ $cm^{-1} sr^{-1}$ for the middle layer at the center frequency. Data from this phantom has been reported in the previous publication \cite{nam2012comparison}.  

\begin{figure}	
	
	\centering
	\includegraphics[width=0.20\textwidth]{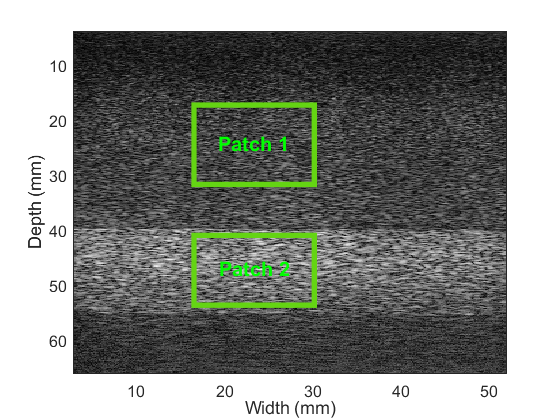}
	\includegraphics[width=0.20\textwidth]{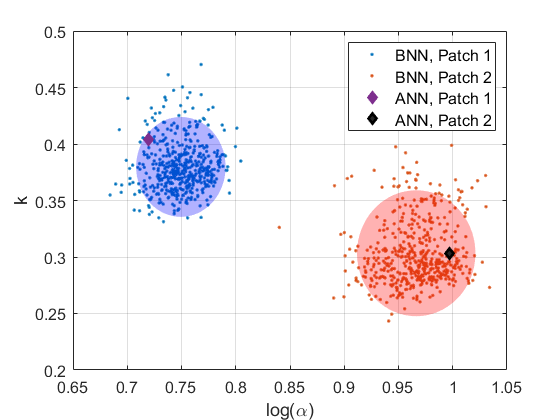}
	\caption{B-mode image of the layered phantom (top) and predictions for the patches specified in the b-mode image (bottom) for BNN and ANN trained on $N_s = 16384$. The shaded areas show the 2 times of the standard devation of the predictions.}
	\label{fig:exp}
\end{figure}

The B-mode image of the phantom is shown in Fig. \ref{fig:exp} (top). Two large patches of size $14.40\times 13.6$ $mm$ (patch 1) and $12.68\times 13.6$ $mm$ are extracted from low and high scatterer concentration layers, respectively. In order to avoid introducing bias, neighbor samples (14 samples in axial and 3 in lateral) are skipped  to reduce the correlation between samples before computing the features. The obtained features are averaged over 12 frames and then given to the networks. The features were passed to the BNN multiple times to acquire different samples of the predicted distribution. The results are shown in Fig. \ref{fig:exp}.

The patch 2 has higher $\alpha$ than the patch 1 which is expected since patch 2 has A higher scatterer concentration. Although the phantom has very low coherent components, the predicted $k$ parameter is discernible. One possible explanation could be the false coherency due to low number of samples \cite{rosado2016analysis}. Comparing the two methods, ANN only provides a single estimate of the parameters while, BNN offers the distribution of the parameters which can be sampled multiple times. By looking closely at the BNN results, it can be observed that the network has a higher uncertainty for patch 2. This has physical interpretation that by increasing the scatterer number density, the estimation would be more difficult and a higher uncertainty is obtained.         

The exact value of $\alpha$ is not known for the phantom but the ratio of high to low scatterer density is close to the ratio of their corresponding backscattering coefficients which is known. The mean value $\pm$ standard deviation of the BNN prediction of $log_{10}(\alpha)$ for the patch 1 is $0.749\pm0.0206$, and it is $0.967\pm0.0273$ for the patch 2. The ratio of backscattering coefficients of patch 2 to patch 1 is $\frac{6.37}{3.52} = 1.81$. The ratio of the predicted $\alpha$ values is $\frac{10^{0.967\pm0.0273}}{10^{0.749\pm0.0206}} = 1.65\pm0.128$. It can be observed that the ratio of the estimated $\alpha$ values is very close to the ground truth ratio of backscattering coefficients.     

\section{Conclusion}
In this paper, a Bayesian Neural Network (BNN) is proposed to estimate HK-distribution parameters. The method provides the distribution of estimated parameters which can be sampled multiple times to acquire the mean prediction and uncertainty. It is compared with a recent neural network approach using simulation and experimental phantom data.  
\section{Compliance with ethical standards}
This is a numerical and experimental phantom study for which no ethical
approval was required.
\section{Acknowledgments}
We acknowledge the support of the Natural Sciences and Engineering
Research Council of Canada (NSERC), and the report of the phantom data in the previous publication \cite{nam2012comparison} .

\bibliographystyle{IEEEbib}
\bibliography{refs3}

\end{document}